\def\nk{n_{\rm b}}

\def\Pb{P_{\rm b}}

\def\rfr#1{Equation~(\ref{#1})}
\def\rfrs#1#2{Equations~(\ref{#1})~to~(\ref{#2})}

\def\derp#1#2{\rp{\partial{#1}}{\partial{#2}}}
\def\dert#1#2{\frac{{{\textrm{d}}}{#1}}{{{\textrm{d}}}{#2}}}

\def\virg#1{``#1"}

\def\eqi{\begin{equation}}
\def\eqf{\end{equation}}
\def\eqia{\begin{eqnarray}}
\def\eqfa{\end{eqnarray}}

\def\rp#1#2{{#1\over#2}}
\def\lb#1{\label{#1}}

\def\bds#1{\boldsymbol{#1}}

\def\cO{\cos\mathit{\Omega}}
\def\sO{\sin\mathit{\Omega}}

\def\cI{\cos I}
\def\sI{\sin I}

\def\ton#1{\left(#1\right)}
\def\qua#1{\left[#1\right]}
\def\grf#1{\left\{#1\right\}}

\documentclass[onecolumn]{aastex}

\usepackage{hyperref}
\usepackage{booktabs}
\usepackage[table,xcdraw]{xcolor}
\usepackage{multirow}
\usepackage{rotating,tabularx}
\usepackage{float}
\usepackage{amsmath,textgreek,w-greek,wasysym}
\usepackage{amsthm}
\usepackage{amscd,lineno}
\usepackage{amssymb,dsfont}
\usepackage{graphicx,epsfig}
\usepackage{txfonts}
\usepackage{xr-hyper}

\RequirePackage{color}

\newcommand{\emaila}{lorenzo.iorio@libero.it}

\allowdisplaybreaks[1]

\begin{document}

\title{Is it possible to measure new general relativistic third-body effects on the orbit of Mercury with BepiColombo?}

\shortauthors{L. Iorio}

\author{Lorenzo Iorio\altaffilmark{1} }
\affil{Ministero dell'Istruzione, dell'Universit\`{a} e della Ricerca
(M.I.U.R.)-Istruzione
\\ Permanent address for correspondence: Viale Unit\`{a} di Italia 68, 70125, Bari (BA),
Italy}

\email{\emaila}

\begin{abstract}
Recently, Will calculated an additional contribution to the Mercury's precession of the longitude of perihelion $\varpi$  of the order of $\dot\varpi_\textrm{W}\simeq 0.22~\textrm{milliarcseconds~per~century}$ ($\textrm{mas~cty}^{-1}$).
It is partly a direct consequence of certain 1pN third-body accelerations entering the planetary equations of motion,  and partly an indirect, mixed effect due to the simultaneous interplay of the standard 1pN pointlike acceleration of the primary with the Newtonian $N$-body acceleration, to the quadrupole order, in the analytical calculation of the secular perihelion precession with the Gauss equations. We critically discuss the actual measurability of the mixed effects with respect to direct ones. The current uncertainties in either the magnitude of the Sun's angular momentum $S_\odot$ and the orientation of its spin axis ${\bds{\hat{S}}}_\odot$ impact the precessions $\dot\varpi_{J_2^\odot},~\dot\varpi_\textrm{LT}$ induced by the Sun's quadrupole mass moment and angular momentum via the Lense-Thirring effect to a level which makes almost impossible to measure $\dot\varpi_\textrm{W}$, even in the hypothesis that it comes entirely from the aforementioned 1pN third-body accelerations. On the other hand, from the point of view of the Lense-Thirring effect itself, the mismodeled quadrupolar precession $\delta\dot\varpi_{J_2^\odot}$ due to the uncertainties in ${\bds{\hat{S}}}_\odot$ corresponds to a bias of $\simeq 9\%$ of the relativistic one. The resulting simulated mismodeled range and range-rate times series of BepiColombo are at about the per cent level of the nominal gravitomagnetic ones.
\end{abstract}

keywords{
Relativity and gravitation; Experimental studies of gravity; Experimental tests of gravitational theories; ephemerides, almanacs, and calendars;
lunar, planetary, and deep-space probes
}

%

\section{Introduction}
Recently, \citet{Will018} calculated a new general relativistic contribution
\eqi
\dot\varpi_\textrm{W} \simeq 0.22~\textrm{mas~cty}^{-1}\lb{totperi}
\eqf
to the secular precession of the longitude of the perihelion $\varpi$ of Mercury arising from the other planets of our solar system up to Saturn.
A similar scenario, but with the perturbing body moving in an inner orbit with respect to the test particle, was treated in \citet{2012MNRAS.423.3540Y}.
The precession of \rfr{totperi} is, partly, a direct consequence of some post-Newtonian accelerations of order $\mathcal{O}\ton{c^{-2}}$ (1pN) induced by a distant, pointlike body X; see\footnote{Note that the appellative \virg{Cross} in Eq.~(4) of \citet{Will018} may turn out somewhat misleading if taken literally in that it may induce an inattentive reader to believe, at a first superficial reading,  that it refers to a mixing of standard Newtonian and pN accelerations in the analytical calculation of the secular effects through the standard Gauss perturbative scheme \citep{2015IJMPD..2450067I}. Instead, in a broad sense, it simply  points to the presence of both the primary and X in certain pN accelerations. } $\qua{\bds a}_\textrm{Cross}$ in Eq.~(4) of \citet{Will018}. On the other hand, a mixed, indirect contribution, allegedly of the same order of magnitude of the direct ones, comes also from the interplay between the standard Newtonian third-body\footnote{The perturber X was assumed to move in a circular orbit coplanar with the Sun-Mercury orbital plane \citep{Will018}. Also \citet{2012MNRAS.423.3540Y} made the same assumptions.} acceleration, which, to the quadrupole order, is
\eqi
{\bds A}_\textrm{X} = -\rp{\mu_\textrm{X}r}{r_\textrm{X}}\qua{\bds{\hat{r}}-3\ton{\bds{\hat{r}}\bds\cdot{\bds{\hat{r}}}_\textrm{X}}
{\bds{\hat{r}}}_\textrm{X}},\lb{Am3}
\eqf
and the usual 1pN pointlike acceleration due to only the primary's mass
\eqi
{\bds A}^M_\textrm{1pN} = \rp{\mu}{c^2 r^2}\qua{\ton{\rp{4\mu}{r} - v^2}\bds{\hat{r}} + 4\ton{\bds{v}\bds\cdot\bds{\hat{r}}}\bds{v}   }\lb{A1PN}
\eqf
in the perturbative calculation by means of the Gauss equations
inasmuch the same way as in the case of the Newtonian acceleration due to the quadrupole mass moment of the primary and \rfr{A1PN} \citep{2014PhRvD..89d4043W,2015IJMPD..2450067I}.
In particular, the largest contribution
\eqi
\dot\varpi_\textrm{W~max}\simeq 0.16~\textrm{mas~cty}^{-1}\lb{Willperi}
\eqf
to the new precession of \rfr{totperi} is due to the direct and mixed effects which do not depend on the velocity ${\bds{v}}_\textrm{X}$ of the distant perturber. \citet{Will018} did not display the direct and indirect contributions to \rfr{Willperi} separately, so that it is not possible to \textcolor{black}{explicitly} establish the weights of both the effects. Actually, it may have its importance in view of the fact that, as explained below, the mixed effects may be unobservable. The direct acceleration in Eq.~(4) of \citet{Will018} which contains ${\bds{v}}_\textrm{X}$ gives rise to a de Sitter-like precession which is about $0.4$ times smaller than \rfr{Willperi} \citep{Will018}. In Appendix~\ref{esatto}, we offer our contribution by analytically working out the direct precession induced by all the acceleration entering $\qua{\bds{a}}_\textrm{Cross}$ in Eq.~(4) of \citet{Will018} without making any simplifying assumptions concerning the orbital configuration of both the perturbed test particle and the distant pointlike perturber X. For Mercury, we find a total pN third-body perihelion precession induced by the planets from Venus to Saturn which amounts to $0.15~\textrm{mas~cty}^{-1}$, which disagrees with \rfr{totperi}. In particular, the total direct precession due to the first two accelerations entering Eq.~(4) of \citet{Will018} amounts to $0.087~\textrm{mas~cty}^{-1}$ instead of\footnote{According to a personal communication by C.~M. Will  to the author, the rest is due to the indirect, mixed effects.} \rfr{Willperi}.

In view of the fact that the largest 1pN contribution to the Mercury's perihelion precession of
\eqi
\dot\varpi_\textrm{1pN} = \ton{\rp{2 + 2 \gamma - \beta}{3}}\rp{3\nk\mu}{c^2 p} = \ton{\rp{2 + 2 \gamma - \beta}{3}}42.98~\textrm{arsec~cty}^{-1}\lb{GR}
\eqf is rescaled in terms of the PPN parameters $\beta,~\gamma$, which are equal to 1 in general relativity,  \citet{Will018} argues that, since the forthcoming BepiColombo mission is expected to improve out knowledge of $\beta,~\gamma$ to the $10^{-6}$ level \citep{2016Univ....2...21S,2018Icar..301....9I}, then it would be likely possible to measure \rfr{totperi}. Indeed, the resulting theoretical mismodeling in \rfr{GR} would be as little as
\eqi
\delta\dot\varpi_\textrm{GR} \simeq 0.03~\textrm{mas~cty}^{-1}.
\eqf
More specifically, \citet{Will018} in the Abstract writes: \virg{At a few parts in $10^{-6}$ of the leading general relativistic
precession of $42.98$ arcseconds per century, these effects are likely to be detectable by the BepiColombo
mission}. Furthermore, \citet{Will018} at pag. 191101-4 writes: \virg{If BebiColombo can reach a part per million
accuracy in measuring the perihelion advance, [\ldots] it will measure, for the
first time, relativistic effects on Mercury's orbit arising
from the planets that surround it.} Conversely, if one is interested in determining the Sun's quadrupole mass moment and angular momentum through their precessions, \rfr{totperi} would act as a systematic bias on them. \citet{Will018} at pag. 191101-4 writes about his new effects: \virg{[\ldots] their existence and cross-correlations may play a role [\ldots] in measurements of the contributions to Mercury's perihelion advance arising from the solar quadrupole moment and frame dragging that will be carried out using data from BepiColombo}.

In this Communication, we will show that measuring \rfr{totperi}, or our smaller result in Appendix~\ref{esatto}, is unlikely, mainly because of the uncertainties in the magnitude of the Sun's angular momentum entering the gravitomagnetic apsidal rate of change  and in the spatial orientation of the Sun's spin axis  affecting especially the precession  induced by the solar quadrupole mass moment. As a byproduct, our results will be useful in assessing the impact of the latter source of systematic uncertainty on the possible measurement of the Lense-Thirring effect itself with BepiColombo. Finally, our exact calculation of the direct precessions have a general validity, and can be fruitfully applied in several astronomical and astrophysical scenarios like, e.g., exoplanets or the stellar system orbiting the supermassive black hole in the Galactic Center characterized by arbitrary eccentricities and inclinations.
\section{Our analysis}
As a general remark, we note that the indirect, mixed effects, which arise from the simultaneous interplay of at least two accelerations $\textrm{A,~B}$ in the calculation of the averaged precessions of the Keplerian orbital elements with the Gauss equations \citep{2014PhRvD..89d4043W,2015IJMPD..2450067I}, are likely undetectable in practical data reductions. Indeed, as far as our case is concerned in which A is, say, \rfr{A1PN} and B is \rfr{Am3}, data analysts of virtually all groups scattered around the world routinely model the Newtonian $N$-body interactions and the 1PN pointlike acceleration due to the primary to
the best of our current knowledge of the parameters entering them which, of
course, is necessarily imperfect. Thus, the actual output of data reductions like residuals of, say,
ranges, range-rates, etc. would not show the
indirect, mixed effects in full. They could only contain negligible signatures, if any,
due to the mismodeling in the planetary masses and in the PPN parameters $\beta,\gamma$ in terms of which the 1PN point particle acceleration is expressed.
Instead, at least in principle, the observables' residuals should fully
display the direct effects (unless they have been somewhat removed
in the estimation of, say, the initial state vectors) induced by some new accelerations, like those of $[\bds a]_\textrm{Cross}$ in Eq.~(4) of \citet{Will018} which, perhaps, may still not be included in the dynamical
models fit to the observations by some groups. Otherwise, one should not model
both \rfr{A1PN} and \rfr{Am3} at all, and subtract their theoretically
computed signals from the resulting huge residuals. It does not
seem certainly viable. Even from the point
of view of a covariance analysis, while it would be possible, in principle, to
explicitly solve for and estimate dedicated scaling parameter(s) accounting for every single acceleration entering the equations of motion, this could
not be done for the indirect, mixed effects.
In the following analysis, we will treat \rfr{totperi} as if it were a potentially measurable effect, irrespectively of its origin.

In addition to the well known 1PN pointlike precession of \rfr{GR} due to solely the primary's mass, there are other two further effects affecting the perihelion of Mercury which should be regarded as serious sources of potential systematic uncertainties: they are due to the first even zonal harmonic $J_2^\odot$ of the multipolar expansion of the Sun's Newtonian gravitational potential, and the general relativistic gravitomagnetic field of the Sun induced by its angular momentum ${\bds{S}}_\odot$. Their precessions depend not only on the size of $J_2^\odot,~S_\odot$, but also on the orientation of the Sun's spin axis ${\bds{\hat{S}}}_\odot$ in space which must enter the error budget as well. Their exact expressions, valid in any coordinate system and for arbitrary orbital configurations, are \citep{2011PhRvD..84l4001I}
\begin{align}
\dot\varpi_{J_2} &= -\rp{3\nk R^2 J_2}{4p^2}\grf{2\qua{-1 + \ton{\bds{\hat{S}}\bds\cdot\bds{\hat{\mathrm{m}}}}\ton{\bds{\hat{S}}\bds\cdot\bds{\hat{\mathrm{n}}}}\ton{1-\cot I}} +3\qua{\ton{\bds{\hat{S}}\bds\cdot\bds{\hat{\mathrm{m}}}}^2 + \ton{\bds{\hat{S}}\bds\cdot\bds{\hat{\mathrm{l}}}}^2} },\lb{dvarpidtJ2} \\ \nonumber \\
\dot\varpi_\textrm{LT} & = -\rp{2GS}{c^2 a^3\ton{1-e^2}^{3/2}}\bds{\hat{S}}\bds\cdot\qua{2~\bds{\hat{\mathrm{n}}} + \ton{\cot I-\csc I}\bds{\hat{\mathrm{m}}}}\lb{dvarpidtLT}.
\end{align}
The Sun's quadrupole mass moment and angular momentum are currently known to the level of accuracy listed in Table \ref{tab0} along with the nominal values of the precessions of \rfrs{dvarpidtJ2}{dvarpidtLT}. It can be noted that, if, on the one hand, it could be hoped that the expected determinations of $J_2^\odot$ by BepiColombo to the $\simeq 10^{-10}$ level \citep{2007PhRvD..75b2001A,2016Univ....2...21S,2018Icar..301....9I} may be accurate enough to make \rfr{totperi} at least larger than the mismodelled $J_2^\odot$-induced precession, on the other hand, a lingering $\simeq 6\%$ uncertainty in $S_\odot$ would imply an a priori theoretical uncertainty in the Lense-Thirring precession of \rfr{dvarpidtLT} as large as $0.13~\textrm{mas~cty}$ corresponding to $\simeq 58\%$ of \rfr{totperi} and $\simeq 86\%$ of our result in Table~\ref{tab1}.

\begin{figure*}
\centering{
\vbox{
\begin{tabular}{c}
\epsfig{file = 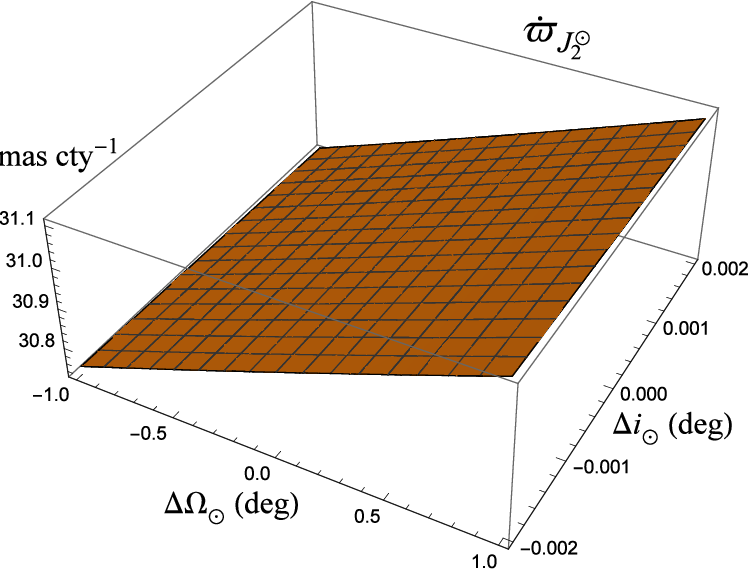,width=0.70\linewidth,clip=}\\
\end{tabular}
}
}
\caption{Plot of $\dot\varpi_{J_2^\odot}\ton{\Delta\mathrm{\Omega}_\odot,~\Delta i_\odot}$, with the Sun's spin axis $\bds{\hat{S}}_\odot$ parameterized in terms of the Carrington elements $\mathrm{\Omega}_\odot,~i_\odot$,  as a function of $\Delta\mathrm{\Omega}_\odot,~\Delta i_\odot$ allowed to vary within $\mp 1~\textrm{deg},~\mp 0.002~\textrm{deg}$ \citep{2005ApJ...621L.153B}, respectively. As a model of the $J_2^\odot$-induced precession of Mercury, \rfr{dvarpidtJ2} was used along with  $J_2^\odot = 2.295\times 10^{-7}$ \citep{2017NSTIM.108.....V}, and  $\Omega_\odot = 73.5~\textrm{deg},~i_\odot = 7.155~\textrm{deg}$ \citep{2005ApJ...621L.153B}. The full range of variation amounts to about $\Delta\dot\varpi_{J_2^\odot}\simeq 0.35~\textrm{mas~cty}^{-1}$. Cfr. with Figure~\ref{fig1}. It is just twice the error calculated in \rfr{erroJ2}.}\label{fig0}
\end{figure*}

\begin{figure*}
\centering{
\vbox{
\begin{tabular}{c}
\epsfig{file = 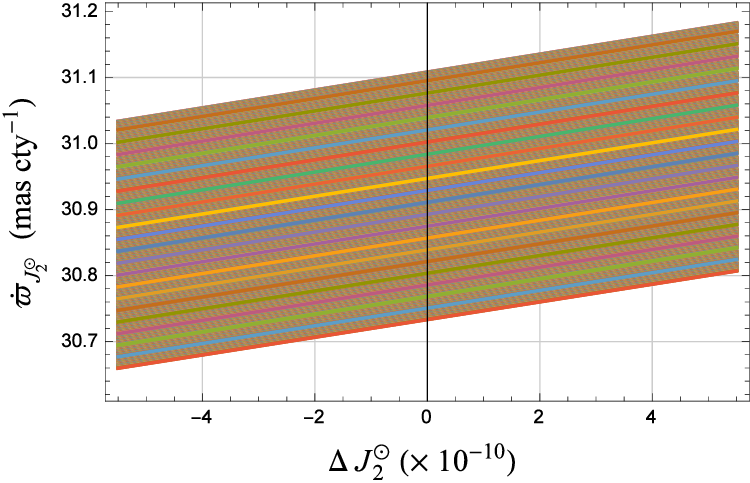,width=0.70\linewidth,clip=}\\
\end{tabular}
}
}
\caption{Family of parametric plots of $\dot\varpi_{J_2^\odot}\ton{\Delta J_2^\odot;~\Delta\mathrm{\Omega}_\odot,~\Delta i_\odot}$, with the Sun's spin axis $\bds{\hat{S}}_\odot$ expressed in terms of the Carrington elements $\mathrm{\Omega}_\odot,~i_\odot$, as a function of $\Delta J_2^\odot$ allowed to vary within  $\mp 5.5\times 10^{-10}$ \citep{2018Icar..301....9I}. As a model of the $J_2^\odot$-induced precession of Mercury, \rfr{dvarpidtJ2} was used along with the reference values  $J_2^\odot = 2.295\times 10^{-7}$ \citep{2017NSTIM.108.....V}, and  $\Omega_\odot = 73.5~\textrm{deg},~i_\odot = 7.155~\textrm{deg}$ \citep{2005ApJ...621L.153B}. Each curve corresponds to given values of $\Delta\mathrm{\Omega}_\odot,~\Delta i_\odot$ within $\mp 1~\textrm{deg},~\mp 0.002~\textrm{deg}$ \citep{2005ApJ...621L.153B}, respectively. For fixed values of $\Delta J_2^\odot$, the full range of variation amounts to about $\Delta\dot\varpi_{J_2^\odot}\simeq 0.35~\textrm{mas~cty}^{-1}$, in agreement with Figure~\ref{fig0}. It is just twice the error calculated in \rfr{erroJ2}.}\label{fig1}
\end{figure*}

\begin{figure*}
\centering{
\vbox{
\begin{tabular}{cc}
\epsfig{file = 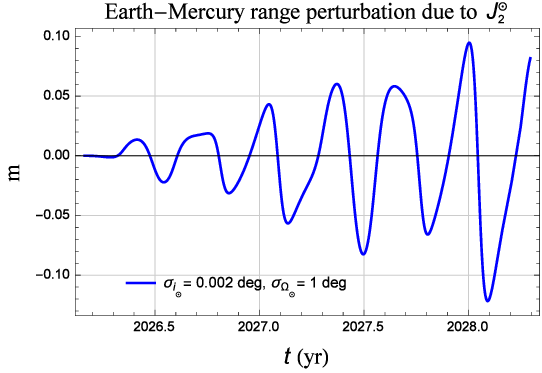,width=0.47\linewidth,clip=}&\epsfig{file = 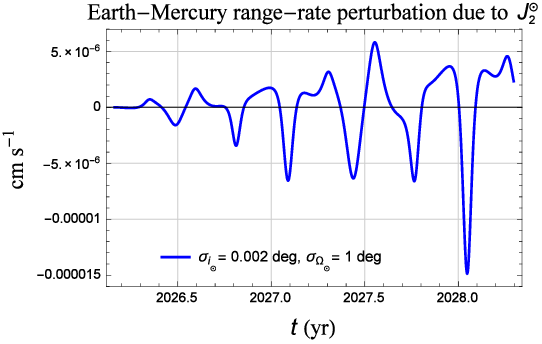,width=0.47\linewidth,clip=}\\
\epsfig{file = 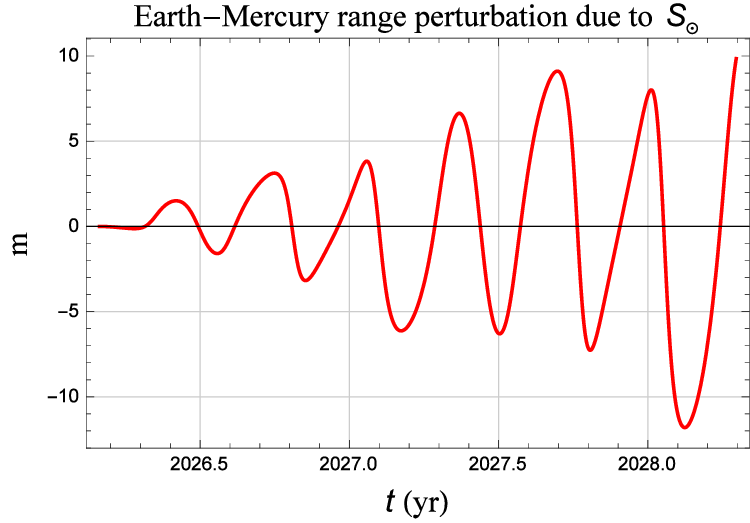,width=0.47\linewidth,clip=}&\epsfig{file = 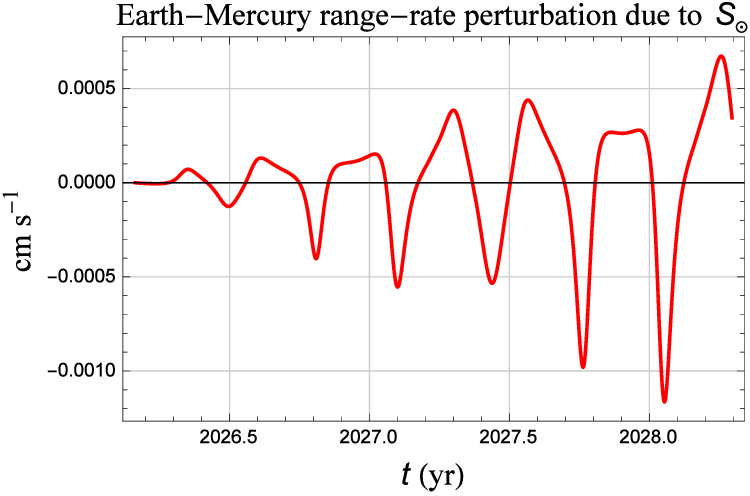,width=0.47\linewidth,clip=}\\
\end{tabular}
}
}
\caption{Upper row (blue): mismodelled Earth-Mercury range (in m) and range-rate (cm s$^{-1}$) $J_2^\odot$-induced perturbations due to the uncertainties $\upsigma_{\Omega_\odot},~\upsigma_{i_\odot}$ in the Carrington elements $\Omega_\odot,~i_\odot$ of the Sun's spin axis ${\bds{\hat{S}}}_\odot$ as in \citet{2005ApJ...621L.153B} during the expected extended mission of BepiColombo from 2026 March 14 to 2028 May 1. Lower row (red): nominal Earth-Mercury range and range-rate perturbations due to the Sun's angular momentum $S_\odot$ through the Lense-Thirring effect during the same temporal interval. A coordinate system with the mean ecliptic at the epoch J2000.0 as fundamental reference $\grf{x,~y}$ plane was assumed. The initial values of the Earth and Mercury osculating orbital elements were retrieved from the Web-Interface HORIZONS maintained by the JPL, NASA. For the nominal values of the Sun's physical parameters used, see Table~\ref{tab0}. }\label{fig2}
\end{figure*}

As announced before, also the current uncertainties  in the Carrington elements parameterizing ${\bds{\hat{S}}}_\odot$ play a crucial role in view of the resulting mismodeling in \rfr{dvarpidtJ2}.
Indeed, a standard Root-Sum-Square (RSS) calculation of the error in $\dot\varpi_{J_2^\odot}$ due to the uncertainties in $i_\odot,~\Omega_\odot$, treated as two independent variables, yields
\eqi
\delta\dot\varpi_{J_2^\odot}<\sqrt{\ton{\derp{\dot\varpi_{J_2}}{\Omega_\odot}}^2\upsigma^2_{\Omega_\odot} + \ton{\derp{\dot\varpi_{J_2}}{i_\odot}}^2\upsigma^2_{i_\odot}} = 0.18~\textrm{mas~cty}^{-1}\lb{erroJ2}
\eqf
Furthermore, Figs.~\ref{fig0}~to~\ref{fig1} straightforwardly depict \rfr{dvarpidtJ2} as function of $J_2^\odot,~i_\odot,~\Omega_\odot$ as independent variables allowed to vary within their ranges of assumed uncertainties \citep{2018Icar..301....9I,2005ApJ...621L.153B}. Their full range of variation is about twice \rfr{erroJ2}.
Instead, as shown by
\eqi
\delta\dot\varpi_\textrm{LT}<\sqrt{\ton{\derp{\dot\varpi_\textrm{LT}}{\Omega_\odot}}^2\upsigma^2_{\Omega_\odot} + \ton{\derp{\dot\varpi_\textrm{LT}}{i_\odot}}^2\upsigma^2_{i_\odot}} = 3\times 10^{-4}~\textrm{mas~cty}^{-1},\lb{erroLT}
\eqf
the Lense-Thirring precession is not significantly impacted by the uncertainty in the Sun's spin axis orientation.
From the point of view of a possible measurement of the Lense-Thirring effect, \rfr{erroJ2} corresponds to a $9\%$ uncertainty in the gravitomagnetic precession. Fig.~\ref{fig2} shows the impact of the uncertainties in the Carrington elements on the direct BepiColombo observables, i.e. range and range-rate. It can be noticed that the resulting mismodeled signatures amount to $\simeq 1-1.5\%$ of the nominal Lense-Thirring ones.
\citet{2018arXiv180402996S}, with dedicated covariance analyses performed with simulated data of BepiColombo, detailed the practical difficulty of satisfactorily separating $J_2^\odot$ from $S_\odot$, and the impact of $S_\odot$ itself in estimating of $J_2^\odot$ in various scenarios.
\section{Conclusions}
The overall post-Newtonian third-body precession of the longitude of the perihelion of Mercury recently calculated by \citet{Will018} amounts to $\dot\varpi_\textrm{W} \simeq 0.22~\textrm{mas~cty}^{-1}$; according to \citet{Will018}, it should be measurable by the forthcoming BepiColombo mission. If, on the one hand, a determination of $J_2^\odot$ at the $\simeq 5\times 10^{-10}$ level, expected from BepiColombo, may reduce the mismodeling in the quadrupolar perihelion precession of Mercury down to $\delta\dot\varpi_{J_2^\odot}\simeq 35\%~\dot\varpi_\textrm{W} \simeq  4\%~\dot\varpi_\textrm{LT}$, on the other hand, the uncertainties in ${\bds{\hat{S}}}_\odot$ would yield $\delta\dot\varpi_{J_2^\odot}\simeq 81\%~\dot\varpi_\textrm{W} = 9\%~\dot\varpi_\textrm{LT}$. Furthermore, the current $\simeq 6\%$ uncertainty in $S_\odot$ would cause a further bias as large as $\delta\dot\varpi_\textrm{LT}\simeq 58\%~\dot\varpi_\textrm{W}$.
It seems that the indirect contributions to $\dot\varpi_\textrm{W}$ arising from the mixing of the Newtonian $N$-body term with the 1pN pointlike acceleration of the Sun in the perturbative analytical calculation, which may not be measurable, amounts to about $0.07~\textrm{mas~cty}^{-1}$. Indeed, our \textcolor{black}{own} calculation returns $0.15~\textrm{mas~cty}^{-1}$ for the total direct post-Newtonian perihelion precession of Mercury induced by the other planets from Venus to Saturn, making, thus, even more pessimistic the perspective of measuring it. The simulated Earth-Mercury range and range-rate time series due to the imperfect knowledge of ${\bds{\hat{S}}}_\odot$ are about at a per cent level of the nominal Lense-Thirring signatures. Finally, we note that our exact calculation for such kind of general relativistic precessions are valid for any orbital configuration of both the test particle and the third body. Thus, they can be applied also to other astronomical and astrophysical natural laboratories characterized by large eccentricities and inclinations like, e.g., several exoplanetary systems and the stars orbiting the supermassive black hole in Sgr A$^{\ast}$ in which the coplanarity condition is not fulfilled.
\section*{Acknowledgements}
I am grateful to C.~M. Will for useful communications.
\appendix
\section{Notations and definitions}\lb{appen}
Here, basic notations and definitions used in the text are presented \citep{1991ercm.book.....B,Nobilibook87,1989racm.book.....S,2003ASSL..293.....B,2005ApJ...621L.153B}
\begin{description}
\item[] $G:$ Newtonian constant of gravitation
\item[] $c:$ speed of light in vacuum
\item[] $m_\textrm{X}:$ mass of the distant pointlike perturber X
\item[] $\mu_\textrm{X}\doteq Gm_\textrm{X}:$ gravitational parameter of the distant pointlike perturber X
\item[] $r_\textrm{X}:$ distance of the distant pointlike perturber X from the primary
\item[] ${\bds{\hat{r}}}_\textrm{X}:$ unit vector of the position vector of the distant pointlike perturber X
\item[] ${\bds{v}}_\textrm{X}:$ velocity vector of the distant pointlike perturber X
\item[] $M:$ mass of the primary
\item[] $\mu\doteq GM:$ gravitational parameter of the primary
\item[] $R:$ equatorial radius of the primary
\item[] $J_2:$ dimensionless quadrupole mass moment of the primary
\item[] $S:$ magnitude of the angular momentum of the primary
\item[] $\bds{\hat{S}}:$ unit vector of the spin axis of the primary
\item[] $\mathrm{\Omega}_\odot:$ longitude of the ascending node of the Sun's equatorial plane with respect to the Vernal Equinox $\aries$ along the Ecliptic. One of the Carrington elements
\item[] $i_\odot:$ inclination of the Sun's equatorial plane to the plane of the Ecliptic. One of the Carrington elements
\item[] ${\bds{\hat{S}}}_\odot = \grf{\sin i_\odot\sin\mathrm{\Omega}_\odot,~-\sin i_\odot\cos\mathrm{\Omega}_\odot,~ \cos i_\odot}:$ Sun's spin axis unit vector in terms of the Carrington elements
\item[] $r:$ distance of the test particle from the primary
\item[] $\bds{\hat{r}}:$ unit vector of the position vector of the test particle
\item[] $\bds{v}:$ velocity vector of the test particle
\item[] $a:$  semimajor axis
\item[] $\nk \doteq \sqrt{\mu a^{-3}}:$   Keplerian mean motion
\item[] $e:$  eccentricity
\item[] $p\doteq a(1-e^2):$  semilatus rectum
\item[] $I:$  inclination of the orbital plane to the reference $\grf{x,~y}$ plane adopted
\item[] $\mathit{\Omega}:$  longitude of the ascending node
\item[] $\bds{\hat{\mathrm{l}}}\doteq\grf{\cO,~\sO,~0}:$ unit vector directed along the line of the nodes toward the ascending node
\item[] $\bds{\hat{\mathrm{m}}}\doteq\grf{-\cI\sO,~\cI\cO,~\sI}:$ unit vector directed transversely to the line of the nodes in the orbital plane
\item[] $\bds{\hat{\mathrm{n}}}\doteq\grf{\sI\sO,~-\sI\cO,~ \cI}:$ unit vector of the orbital angular momentum
\item[] $\omega:$  argument of pericenter
\item[] $\varpi\doteq \mathit{\Omega} + \omega:$ longitude of pericenter
\end{description}
\section{Exact calculation of the direct perihelion precession}\lb{esatto}
The first line of Eq.~(4) of \citet{Will018} returns the following 1pN acceleration of order $\mathcal{O}\ton{G^2}$
\eqi
{\bds A}_{G^2} = \rp{2\mu\mu_\textrm{X}}{c^2 r_\textrm{X}^3}\qua{ \bds{\hat{r}}  - 6 \ton{\bds{\hat{r}} \bds\cdot {\bds{\hat{r}}}_\textrm{X}  }{\bds{\hat{r}}}_\textrm{X} + 3\ton{\bds{\hat{r}} \bds\cdot{\bds{\hat{r}}}_\textrm{X} }^2  \bds{\hat{r}}  }.\lb{AG2}
\eqf
We were successful in obtaining an exact expression for the doubly-averaged perihelion precession induced by \rfr{AG2} without any  a-priori simplifying assumption on the orbital geometries of both the perturbed test particle and the distant pointlike perturber. Nonetheless, it is far too cumbersome to be explicitly displayed here; thus, we show it only to the zeroth order in the eccentricity $e$.
It reads
\begin{align}
\dert\varpi t \nonumber & = -\rp{\mu_\textrm{X}\sqrt{\mu a}}{16 c^2 a^3_\textrm{X}\ton{1 - e^2_\textrm{X}}^{3/2}}\grf{
1 + 48 \ton{-1 + \cos I}\cos I \ton{\cos^2 I_\textrm{X} -\cos^2\Delta\Omega \sin^2 I_\textrm{X}} + \right.\\ \nonumber \\
\nonumber & + \left. 3 \ton{2 \cos 2 \Delta\Omega \sin^2 I + \cos 2 I \ton{1 + \cos 2 I_\textrm{X} \ton{3 + \cos 2 \Delta\Omega - 9 \cos 2\omega} + \right.\right.\right.\\ \nonumber \\
\nonumber & + \left.\left. \left. 6 \cos 2 \Delta\Omega \cos 2\omega \sin^2 I_\textrm{X}} + \cos 2 I_\textrm{X} \ton{9 \cos 2\omega +
2 \sin^2\Delta\Omega} + \right.\right.\\ \nonumber \\
\nonumber & + \left.\left. 2 \ton{3\cos 2\omega \ton{\sin^2 I + 3 \cos 2 \Delta\Omega \sin^2 I_\textrm{X}} + 4 \ton{3 \ton{\sin I \sin 2 I_\textrm{X}\sin\Delta\Omega -\right.\right.\right.\right.\right.\\ \nonumber \\
\nonumber & + \left.\left.\left.\left.\left. \cos I \sin^2 I_\textrm{X} \sin 2\Delta\Omega} \sin 2\omega + \cos\Delta\Omega \sin 2 I_\textrm{X}\ton{-2 \sin I + 3 \sin 2 I \sin^2\omega +
\tan\ton{\rp{I}{2}}}}}}} + \\ \nonumber \\
& + \mathcal{O}\ton{e^2}.\lb{mega}
\end{align}
If the hypothesis of circularity and coplanarity with the test particle is assumed for the orbit of X \cite{Will018}, the exact precession yields a shift per orbit
\eqi
\Delta\varpi = -\rp{2\uppi \mu_\textrm{X}a^2 \sqrt{1-e^2}}{c^2a^3_\textrm{X}}.\lb{shift1}
\eqf

The second line of Eq.~(4) of \citet{Will018} yields the following 1pN acceleration of order $\mathcal{O}\ton{G}$
\eqi
{\bds A}_{G} = \rp{\mu_\textrm{X} r}{c^2 r^3_\textrm{X}}\grf{ 4\bds{v}\qua{\ton{ \bds{v}\bds\cdot \bds{\hat{r}} } - 3\ton{ \bds{\hat{r}}\bds\cdot{\bds{\hat{r}}}_\textrm{X} }\ton{ \bds{v}\bds\cdot{\bds{\hat{r}}}_\textrm{X} }  } - v^2\qua{\bds{\hat{r}} -3\ton{\bds{\hat{r}}\bds\cdot {\bds{\hat{r}}}_\textrm{X}}{\bds{\hat{r}}}_\textrm{X}  } }.\lb{AG1}
\eqf
We were able to calculate its exact, doubly averaged perihelion precession without a priori simplifying assumptions on $e,~I,~\Omega,~\omega,~e_\textrm{X},~I_\textrm{X},~\Omega_\textrm{X},~\omega_\textrm{X}$. Unfortunately, it is particularly cumbersome, and cannot be explicitly displayed here. An important feature of it is that, for arbitrary orbital configurations, the precession due to \rfr{AG1} is not defined for $e\rightarrow0$ since it contains terms of order $\mathcal{O}\ton{e^{-k}},~k=2,~4$. By expanding it in powers of $e$, we have
\begin{align}
\dert\varpi{t} \nonumber &= \rp{9\sqrt{\mu a}\mu_\textrm{X}\ton{1 - 2e^2}}{4c^2 e^4 a_\textrm{X}^3\ton{1 - e_\textrm{X}^2}^{3/2}}\grf{ \cos 2\omega \qua{(1 + 3 \cos 2I_\textrm{X}) \sin^2 I + \right.\right.\\ \nonumber \\
\nonumber &+\left.\left. \ton{3 + \cos 2I}\sin^2 I_\textrm{X}\cos 2\Delta\Omega - 2 \sin 2I \sin 2I_\textrm{X}\cos\Delta\Omega } + \right.\\ \nonumber \\
&+\left. 4\sin 2\omega\ton{-\cos I \sin^2 I_\textrm{X} \sin 2\Delta\Omega + \sin I \sin 2I_\textrm{X} \sin\Delta\Omega}  }+\mathcal{O}\ton{e^0}.\lb{omegae4}
\end{align}
By assuming $e_\textrm{X}=0,~I = I_\textrm{X},~\Omega = \Omega_\textrm{X}$ \citep{Will018}, the resulting full shift per revolution of the test particle turns out to be
\eqi
\Delta\varpi = -\rp{7\uppi \mu_\textrm{X}a^2 \sqrt{1-e^2}}{2c^2a^3_\textrm{X}}.\lb{shift2}
\eqf
In the case of Mercury, the discrepancy between the full precession and the coplanarity-based approximated one, from which  \rfr{shift2} was derived, amounts to $\simeq -0.03~\textrm{mas~cty}^{-1}$ for X=Venus.

The third line of Eq.~(4) of \citet{Will018} provides us with the following 1pN \virg{gravitomagnetic} acceleration of order $\mathcal{O}\ton{G}$ due to the velocity ${\bds{v}}_\textrm{X}$ of the third body
\eqi
{\bds A}_{v_\textrm{X}} = -\rp{\mu_\textrm{X}}{c^2 r_\textrm{X}^2}\qua{4\bds{v}\bds\times\ton{ {\bds{\hat{r}}}_\textrm{X}\bds\times{\bds{v}}_\textrm{X} } - 3\ton{ {\bds{\hat{r}}}_\textrm{X} \bds\cdot{\bds{v}}_\textrm{X} }\bds{v}}.\lb{AGM}
\eqf
Its exact, doubly averaged perihelion precession turns out to be
\eqi
\dert\varpi t = \rp{2\mu_\textrm{X}\nk^\textrm{X}}{c^2 a_\textrm{X}\ton{1-e^2_\textrm{X}}}\qua{\cos I_\textrm{X} + \sin I_\textrm{X}\tan\ton{\rp{I}{2}}\cos\Delta\Omega }.\lb{GMfull}
\eqf
For $\Delta\Omega=0,~I=I_\textrm{X},~e_\textrm{X}=0$, \rfr{GMfull} agrees with the precession which can be inferred from the fourth term of Eq.~(1) of \citet{Will018} by taking the ratio of it to the orbital period $\Pb$ of the perturbed test particle. The numerical discrepancy between \rfr{GMfull} and the approximated expression by Will is negligible; indeed, in the case of Mercury perturbed by Venus, they differ by just $2\times 10^{-5}~\textrm{mas~cty}^{-1}$ yielding both $\dot\varpi = 0.014~\textrm{mas~cty}^{-1}$. The total contribution of all planets from Venus to Saturn to the Mercury's precession of \rfr{GMfull} amounts to $\dot\varpi = 0.06~\textrm{mas~cty}^{-1}$.

See Table~\ref{tab1} for a detailed overview of the contributions of the planets from Venus to Saturn to the Mercury's direct 1pN third-body perihelion precession.
It can be noted that, while the total \virg{gravitomagnetic} effect arising from the third line of Eq.~(4) of \citet{Will018} agrees with the results by \citet{Will018} himself, our total precession due to the first two lines of Eq.~(4) of \citet{Will018} is about half than that claimed by \citet{Will018}. Such a discrepancy seems to be attributable to the indirect, mixed effects\textcolor{black}{, not calculated here}.
\section{Tables and Figures}
\begin{table*}
\caption{Relevant Sun's physical parameters along with the most recent uncertainties for some of them appeared in the literature, and nominal quadrupolar and Lense-Thirring perihelion precessions for Mercury. As far as $S_\odot$is concerned, the values quoted for its size and uncertainty were obtained by calculating the mean and the standard deviation of the figures quoted in Table 1 of  \citet{2012SoPh..281..815I}.}
\label{tab0}
\centering
\begin{tabular}{ll}
\noalign{\smallskip}
\hline
Sun's physical parameters & Value \\
\hline
$\mu_\odot$ \citep{2016AJ....152...41P}& $1.3271244\times 10^{20}~\textrm{m}^3~\textrm{s}^{-2}$\\
$R_\odot$ \citep{2016AJ....152...41P} & $6.957\times 10^8~\textrm{m}$\\
$\Omega_\odot$ \citep{2005ApJ...621L.153B} & $73.5\pm 1~\textrm{deg}$\\
$i_\odot$ \citep{2005ApJ...621L.153B} & $7.155\pm 0.002~\textrm{deg}$\\
$J_2^\odot$ \citep{2017NSTIM.108.....V} & $2.295\times 10^{-7}$\\
$\upsigma_{J_2^\odot}$ \citep{2017AJ....153..121P} & $9\times 10^{-9}$\\
$\upsigma_{J_2^\odot}$ \citep{2018NatureG} & $2.2\times 10^{-9}$\\
$\upsigma_{J_2^\odot}$ \citep{2017NSTIM.108.....V} & $1\times 10^{-9}$\\
$\upsigma_{J_2^\odot}$ \citep{2018Icar..301....9I} & $5.5\times 10^{-10}$\\
$\upsigma_{J_2^\odot}$ \citep{2016Univ....2...21S} & $4.1\times 10^{-10}$\\
$S_\odot$ \citep{2012SoPh..281..815I} & $192.0\times 10^{39}~\textrm{kg~m}^2~\textrm{s}^{-1}$ \\
$\upsigma_{S_\odot}$ \citep{2012SoPh..281..815I} & $12.0\times 10^{39}~\textrm{kg~m}^2~\textrm{s}^{-1}$\\
$\dot\varpi_{J_2^\odot}$ & $31~\textrm{mas~cty}^{-1}$\\
$\dot\varpi_\textrm{LT}$ & $-2~\textrm{mas~cty}^{-1}$\\
\hline
\end{tabular}
\end{table*}
\begin{table*}
\caption{Doubly averaged 1pN third-body perihelion precessions of Mercury, in mas cty$^{-1}$, induced by Venus, Earth, Mars, Jupiter, Saturn via \rfrs{AG2}{AGM}. The resulting total precession amounts to $0.15~\textrm{mas~cty}^{-1}$; in particular, \rfrs{AG2}{AG1} yield a combined overall precession of $0.087~\textrm{mas~cty}^{-1}$, contrary to $0.16~\textrm{mas~cty}^{-1}$ claimed by \citet{Will018}. The discrepancy seems to be due to the indirect, mixed effects\textcolor{black}{, not worked out  here}.}\lb{tab1}
\begin{center}
\begin{tabular}{|r r r r|}
  \hline
   & \rfr{AG2} ($\textrm{mas~cty}^{-1}$) & \rfr{AG1} ($\textrm{mas~cty}^{-1}$) & \rfr{AGM} ($\textrm{mas~cty}^{-1}$) \\
\hline
Venus & $-0.00490$ & $-0.00371$ & $0.01409$ \\
Earth & $-0.00231$ & $0.08183$ & $0.00767$ \\
Mars & $-0.00007$ & $0.00128$ & $0.00029$ \\
Jupiter & $-0.00515$ & $0.05683$ & $0.03967$ \\
Saturn & $-0.00024$ & $-0.00305$ & $0.00260$ \\
\hline
Total & $-0.0127$ & $0.0996$ & $0.0643$\\
\hline
\end{tabular}
\end{center}
\end{table*}
\bibliography{Gclockbib,PXbib,IorioFupeng}{}

\begin{thebibliography}{19}
\expandafter\ifx\csname natexlab\endcsname\relax\def\natexlab#1{#1}\fi

\bibitem[{{Ashby}, {Bender} \& {Wahr}(2007){Ashby}, {Bender}, \&
  {Wahr}}]{2007PhRvD..75b2001A}
{Ashby} N., {Bender} P.~L., {Wahr} J.~M., 2007, Phys. Rev. D, 75, 022001

\bibitem[{{Beck} \& {Giles}(2005)}]{2005ApJ...621L.153B}
{Beck} J.~G., {Giles} P., 2005, Astrophys. J. Lett., 621, L153

\bibitem[{{Bertotti}, {Farinella} \& {Vokrouhlick\'{y}}(2003){Bertotti},
  {Farinella}, \& {Vokrouhlick\'{y}}}]{2003ASSL..293.....B}
{Bertotti} B., {Farinella} P., {Vokrouhlick\'{y}} D., 2003, {Physics of the
  Solar System}. Kluwer, Dordrecht

\bibitem[{{Brumberg}(1991)}]{1991ercm.book.....B}
{Brumberg} V.~A., 1991, {Essential Relativistic Celestial Mechanics}. Adam
  Hilger, Bristol

\bibitem[{{Genova} {et~al}\mbox{.}(2018){Genova}, {Mazarico}, {Goossens},
  {Lemoine}, {Neumann}, {Smith}, \& {Zuber}}]{2018NatureG}
{Genova} A., {Mazarico} E., {Goossens} S., {Lemoine} F.~G., {Neumann} G.~A.,
  {Smith} D.~E., {Zuber} M.~T., 2018, Nature Communications, 9, 289

\bibitem[{{Imperi}, {Iess} \& {Mariani}(2018){Imperi}, {Iess}, \&
  {Mariani}}]{2018Icar..301....9I}
{Imperi} L., {Iess} L., {Mariani} M.~J., 2018, Icarus, 301, 9

\bibitem[{{Iorio}(2011)}]{2011PhRvD..84l4001I}
{Iorio} L., 2011, Phys. Rev. D, 84, 124001

\bibitem[{{Iorio}(2012)}]{2012SoPh..281..815I}
{Iorio} L., 2012, Sol. Phys., 281, 815

\bibitem[{{Iorio}(2015)}]{2015IJMPD..2450067I}
{Iorio} L., 2015, Int. J. Mod. Phys. D, 24, 1550067

\bibitem[{{Milani}, {Nobili} \& {Farinella}(1987){Milani}, {Nobili}, \&
  {Farinella}}]{Nobilibook87}
{Milani} A., {Nobili} A., {Farinella} P., 1987, {Non-gravitational
  perturbations and satellite geodesy}. Adam Hilger, Bristol

\bibitem[{{Park} {et~al}\mbox{.}(2017){Park}, {Folkner}, {Konopliv},
  {Williams}, {Smith}, \& {Zuber}}]{2017AJ....153..121P}
{Park} R.~S., {Folkner} W.~M., {Konopliv} A.~S., {Williams} J.~G., {Smith}
  D.~E., {Zuber} M.~T., 2017, AJ, 153, 121

\bibitem[{{Pr{\v s}a} {et~al}\mbox{.}(2016){Pr{\v s}a}, {Harmanec}, {Torres},
  {Mamajek}, {Asplund}, {Capitaine}, {Christensen-Dalsgaard}, {Depagne},
  {Haberreiter}, {Hekker}, {Hilton}, {Kopp}, {Kostov}, {Kurtz}, {Laskar},
  {Mason}, {Milone}, {Montgomery}, {Richards}, {Schmutz}, {Schou}, \&
  {Stewart}}]{2016AJ....152...41P}
{Pr{\v s}a} A. {et~al.}, 2016, Astron. J., 152, 41

\bibitem[{{Schettino} {et~al}\mbox{.}(2018){Schettino}, {Serra}, {Tommei}, \&
  {Milani}}]{2018arXiv180402996S}
{Schettino} G., {Serra} D., {Tommei} G., {Milani} A., 2018, arXiv:1804.02996

\bibitem[{{Schettino} \& {Tommei}(2016)}]{2016Univ....2...21S}
{Schettino} G., {Tommei} G., 2016, Universe, 2, 21

\bibitem[{{Soffel}(1989)}]{1989racm.book.....S}
{Soffel} M.~H., 1989, {Relativity in Astrometry, Celestial Mechanics and
  Geodesy}. Springer-Verlag; Berlin Heidelberg New York

\bibitem[{{Viswanathan} {et~al}\mbox{.}(2017){Viswanathan}, {Fienga},
  {Gastineau}, \& {Laskar}}]{2017NSTIM.108.....V}
{Viswanathan} V., {Fienga} A., {Gastineau} M., {Laskar} J., 2017, Notes
  Scientifiques et Techniques de l'Institut de M\'{e}canique C\'{e}leste, 108

\bibitem[{{Will}(2014)}]{2014PhRvD..89d4043W}
{Will} C.~M., 2014, Phys. Rev. D, 89, 044043

\bibitem[{{Will}(2018)}]{Will018}
{Will} C.~M., 2018, Phys. Rev. Lett., 120

\bibitem[{{Yamada} \& {Asada}(2012)}]{2012MNRAS.423.3540Y}
{Yamada} K., {Asada} H., 2012, MNRAS, 423, 3540

\end{thebibliography}


\end{document}